# Testing of General Relativity with Geodetic VLBI

© Oleg Titov

Geoscience Australia, Australia

The geodetic VLBI technique is capable of measuring the Sun's gravity light deflection from distant radio sources around the whole sky. This light deflection is equivalent to the conventional gravitational delay used for the reduction of geodetic VLBI data. While numerous tests based on a global set of VLBI data have shown that the parameter $\gamma$ of the post-Newtonian approximation is equal to unity with a precision of about 0.02 percent, more detailed analysis reveals some systematic deviations depending on the angular elongation from the Sun. In this paper a limited set of VLBI observations near the Sun were adjusted to obtain the estimate of the parameter gamma free of the elongation angle impact. The parameter $\gamma$ is still found to be close to unity with precision of 0.06 percent, two subsets of VLBI data measured at short and long baselines produce some statistical inconsistency.

1. Introduction

In accordance with the predictions of general relativity the gravitational field of the Sun causes a light deflection of 1.75" at the Solar limb observed by optical facilities. In the case of the Very Long Baseline Interferometry (VLBI) technique the corresponding effect (known as gravitational delay) is traditionally formulated in a term of time delay and modelled as a function of the barycentre distance of two radio telescopes. A dependence on the baseline length and the elongation from the direction to the Solar system barycentre was not presented in the formula for the gravitational delay. Therefore, these two effects have been considered independently despite their common physical origin. A simple transformation between the deflection angle and the gravitational delay was recently developed to show the baseline length and the elongation angle explicitly.

The most accurate estimate of the post-Newtonian approximation parameter $\gamma$ ($\sigma_\gamma = 2 \cdot 10^{-5}$) was obtained from analysis of data done by the Cassini spacecraft [1]. The current accuracy obtained with VLBI $\sigma_\gamma = 1.5 \cdot 10^{-4}$ was obtained from analysis of a few millions of VLBI observations [2] collected and stored by the International VLBI Service (IVS) [3]. However it was found that the accuracy of the parameter $\gamma$ estimate with geodetic VLBI observations is dominated by a small number of observations of radio sources near the Sun. This paper is emphasised on analysis of a limited set of near-Sun observations rather than a global set of all-sky data.

## 2. Basic equations

The conventional equation for the gravitational delay used for the reduction of geodetic VLBI data is given by

$$\tau_{grav} = \frac{(1+\gamma)GM}{c^3} \ln \frac{|\mathbf{r_1}| + (\mathbf{r_1} \cdot \mathbf{s})}{|\mathbf{r_2}| + (\mathbf{r_2} \cdot \mathbf{s})} \tag{1}$$

where $G$ is the gravitational constant, $r_1$ and $r_2$ are geocentirc distances from a body of mass $M$ to both radio telescopes, $s$ is the unit vector in the direction of the radio source, c is the speed of light and $\gamma$ is the parameter of the post-Newtonian (PPN) formalism [4], equal to unity in GR. The positions of astronomical instruments on Earth are referenced to the solar system barycentre, and the measured delay is equal to the terrestrial time (TT) coordinate time interval between two events of the signal arrival at the first and second radio telescopes [5]. This equation is sufficient for a picosecond level of accuracy.

A deviation of the PPN parameter $\gamma$ from unity is estimated to test general relativity. However, apart from the gravitational delay (1) another term including into the VLBI group delay model also comprises the parameter $\gamma$ [5] (eq (11.9)).

$$\tau_{coord} = \frac{(\gamma+1)GM}{c^2 r} \frac{(\mathbf{b} \cdot \mathbf{s})}{c} \tag{2}$$

where **b** is the baseline vector calculated as the difference between the two barycentre radius-vectors of two antennas $\mathbf{b} = \mathbf{r}_2 - \mathbf{r}_1$.

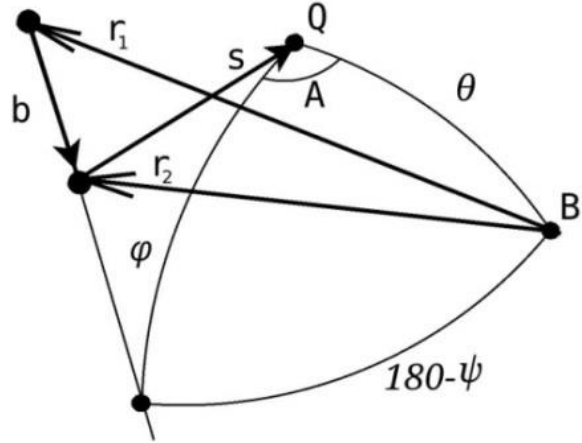

Fig 1: Angles $\psi$, $\varphi$, $\theta$ and $A$, originated by positions of gravitational mass (B), quasar (Q) and baseline vector (b). If the Sun plays the role of the gravitational mass, then the point B is also the position of the Solar system barycentre.

Equation (2) is known to change the scale only and is used to be ignored, so only (1) is used for calculation of the partial derivative.

It was shown that the total effect of general relativity at picosec level accuracy of the PPN approximation for the grazing light is a sum of (1) and (2) $\tau_{GR} = \tau_{grav} + \tau_{coord}$ in a such way that the term (2) vanishes and only physically observable terms are left in the equation [6]

$$\tau_{GR} = \tau_{grav} + \tau_{coord} \approx \frac{(\gamma+1)GMb}{c^3 r_2} \frac{\sin\varphi \sin\theta \cos A}{1-\cos\theta} + \frac{(\gamma+1)GMb^2}{2c^3 r_2^2} \frac{(1-\cos^2\varphi \cos^2\theta)}{(1-\cos\theta)} - \frac{(\gamma+1)GMb^2}{2c^3 r_2^2} \frac{\sin^2\varphi \sin^2\theta \cos^2 A}{(1-\cos\theta)^2} \quad (3)$$

The first term in (4) is linked to the light deflection angle for an arbitrary elongation from the Sun as follows

$$\tau_{GR} = \frac{(\gamma+1)GMb}{c^3 r_2} \frac{\sin\varphi \sin\theta \cos A}{1-\cos\theta} = \alpha_2 \frac{b}{c} \sin\varphi \cos A \quad (4)$$

where $\alpha_2$ is the deflection angle at arbitrary $\theta$ [7, 8]

$$\alpha_2 = \frac{(\gamma+1)GM}{c^2 r_2} \frac{\sin\theta}{1-\cos\theta} \qquad (5)$$

The last two terms in (3) are pertinent only for two-station observational facilities as a standard VLBI baseline, and the additional light deflection is now proportional to the baseline length. The third term is responsible for an incremental deflection along the "source – deflecting body" direction, whereas the second term mimics the baseline parallax effect, i.e. non-zero angle for a radio source observed from two radio telescopes separated by a long baseline. This baseline parallax effect is ignorable for an extragalactic radio source as $\frac{b}{r} \leq 10^{-16}$, however, it is exaggerated by the Solar gravitational field.

Fig. 2 shows the deflection angle variations for radio source 0552+398 at all elongations measured for 10 years (1991–2001) in the IRIS-A VLBI project. The yearly path of the source is a perfect circle with the maximum deflection calculated as in (5) for the minimum elongation that is equal to the ecliptic latitude.

Recalling that $\gamma = 1$ for general relativity, in the small-angle approximation $\tau_{GR}$ is given by

$$\tau_{GR} = \frac{4GM}{c^2 R_2} \frac{b}{c} \sin\varphi \cos A - \frac{2GM}{c^2 R_2} \frac{b}{R_2} \frac{b}{c} \sin^2\varphi + \frac{4GM}{c^2 R_2} \frac{b}{R_2} \frac{b}{c} \sin^2\varphi \cos^2 A \qquad (6)$$

where $R_2 = r_2 \cos\theta_2$.

The first term shows the classic light deflection angle as

$$\alpha = \frac{4GM}{c^2 R_2} \qquad (7)$$

and the last two terms present the baseline-dependent deflection referred to the second station. In accordance with (6), these two minor effects rapidly grow in the vicinity of the Sun (as $\frac{b}{R^2}$) and their sum reaches 12 milliarcsec at a baseline of 10,000 km for the grazing light and 0.03 milliarcsec for an elongation of 5°. For a shorter baseline the magnitudes are proportionally smaller [6].

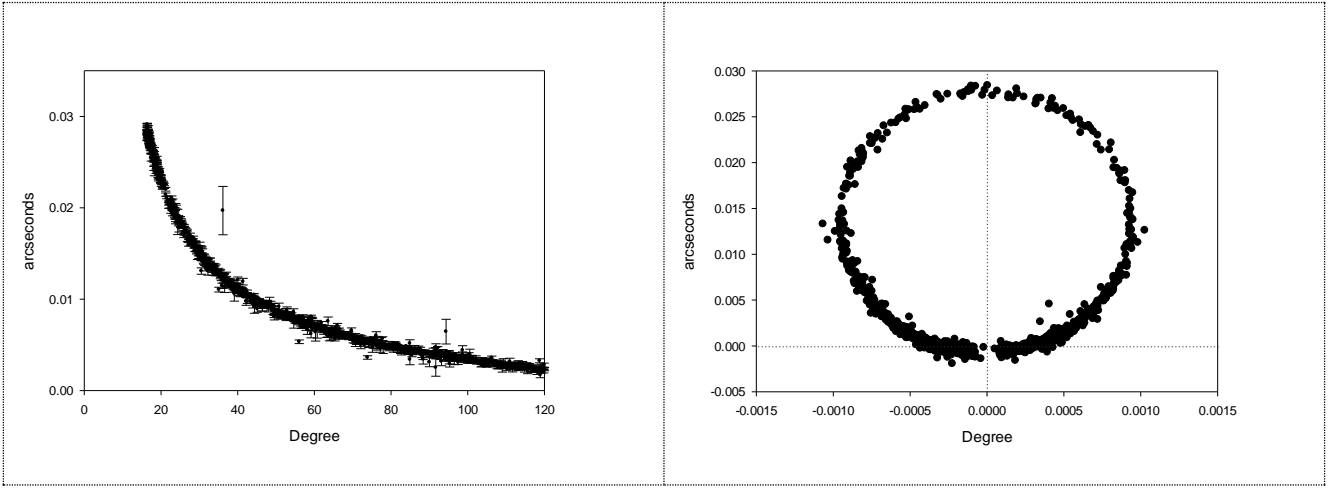

Fig 2. Daily variations of the deflection angle for radio source 0552+398 as a function of the elongation angle (left) and in the sky plane (right). The relativistic effect was not applied for VLBI data reduction

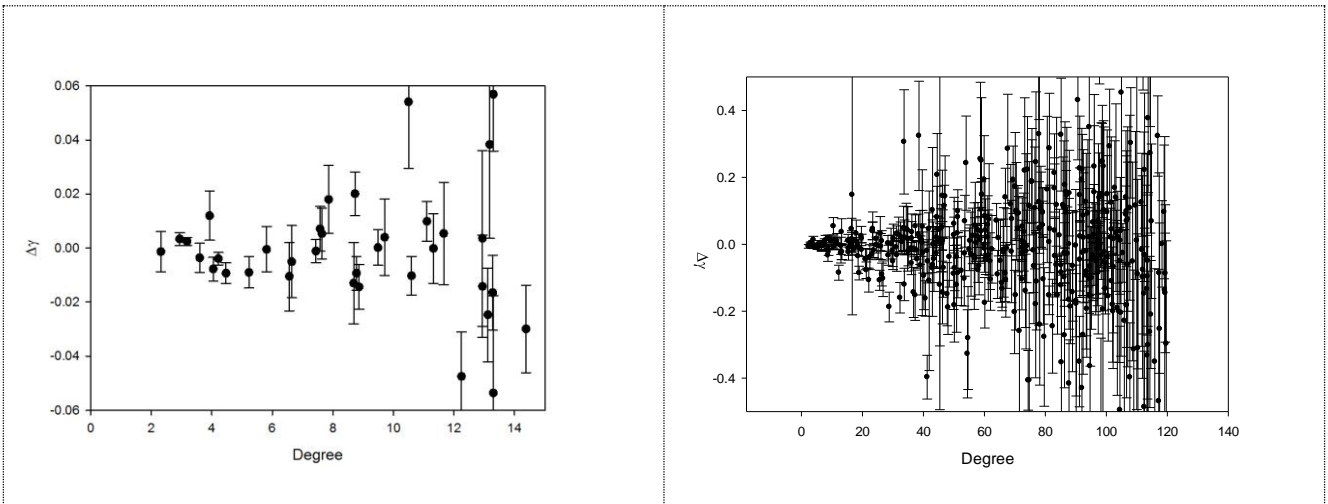

Fig 3. Variations of the daily estimates of the PPN parameter $\gamma$ for radio source 0229+131 in 1991-2001 at the different scale of angle $\theta$. The relativistic effect was applied for VLBI data reduction

## 3. Data analysis

Since the light deflection at large elongation angles are likely to be affected by the numerical and astrometrical factors discussed at the previous section, we focus on the observations of reference radio sources near the Sun to reduce a possible bias or statistical degradation. The legacy VLBI data collected between 1993 and 2002 comprise a large amount of observations of radio sources close to the Sun (at 5° or less), however, there were no observations closer than 15° to the

Sun between 2002 and 2011. As a result, standard IVS observations over that period do not contribute essentially to the improvement of the parameter γ.

In total we processed 53 sessions between 1993 and 2012, these included 58 approaches of radio sources at 5º or less from the Sun. The total number of single observations is 1581. The results of the data analysis are shown on the left panel in Fig. 4. All the single-day estimates of $\gamma$ do not exceed $16 \cdot 10^{-3}$ and no statistical significant deviation from unity was detected. No dependence on the elongation angle or year of experiments was also observed. Nonetheless, it is worth noting that the formal errors for experiments with more than 100 observations (01 May 1996 and 21 Apr 1997) are substantially better. One could conclude that the number of successful scans at even moderate elongation angle is a key factor for future planning of experiments. The solution combining all the close approaches produces the combined estimate of $\gamma$ as $\gamma = (-1.6 \pm 6.5) \cdot 10^{-4}$. This formal error is just 3.5 times worse than the formal error obtained from the recent global solution, which uses about 4.5 million group delays [2].

Also, we estimated the parameter $\gamma$ by splitting the set observations into two parts: at 'short' (≤6000 km) and 'long' baselines (≥6000 km). Surprisingly, the estimates of the correction to γ were not consistent to each other $\gamma = (+12.6 \pm 5.2) \cdot 10^{-4}$ with 820 observations and $\gamma = (-12.0 \pm 4.9) \cdot 10^{-4}$ with 761 observations, correspondingly, as well to the combined solution estimate. Whereas the corrections are not statistically significant at $3\sigma$ level, it is remarkable that the estimates have opposite signs and, moreover, the formal errors are less than for the combined solution formal error in spite of the less number of observations. This could hint either a presence of a hidden systematic term depending on baseline length outside of the conventional relativistic model, or outstanding effects of radio source structure.

To check the hypothesis the estimate of $\gamma$ was obtained with the same set of data for narrower baseline length ranges of 4000 km width (i.e. 1000 – 5000 km, 2000 – 6000 km, etc) (Fig. 4, right panel). Therefore, seven estimates presented on the plot are not independent, although the systematic effect in estimate of $\gamma$ is noticeable. The correction has a tendency to be positive for shorter baselines and negative for longer baselines. The intermediate length baselines (from 5000 to 9000 km) have the best performance among all the estimates $\gamma = (-1.7 \pm 4.2) \cdot 10^{-4}$ based on 902 observations. This accuracy is about 30% better than for the combined solution that included all baselines.

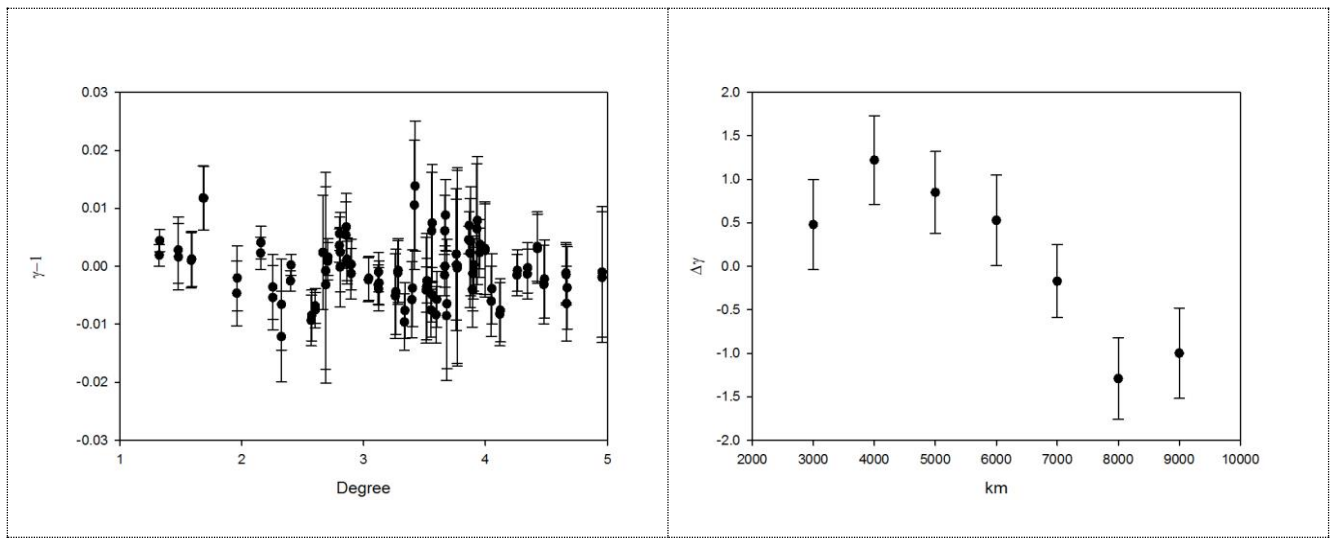

Fig 4. Estimates for correction to the PPN-parameter $\gamma$ from 58 geodetic VLBI sessions with respect to elongation angle (left) and combined estimation of $\gamma$ for different ranges of baseline

**Acknowledgment**